# Going Off-road: Transversal Complexity in Road Networks

David Eppstein

Dept. of Computer Science
University of California, Irvine
Irvine, CA 92697-3435 USA
http://www.ics.uci.edu/~eppstein

Michael T. Goodrich

Dept. of Computer Science
University of California, Irvine
Irvine, CA 92697-3435 USA
http://www.ics.uci.edu/~goodrich

Lowell Trott

Dept. of Computer Science
University of California, Irvine
Irvine, CA 92697-3435 USA
http://www.ics.uci.edu/~ltrott

**Abstract**

A *geometric graph* is a graph embedded in the plane with vertices at points and edges drawn as curves (which are usually straight line segments) between those points. The *average transversal complexity* of a geometric graph is the number of edges of that graph that are crossed by random line or line segment.

In this paper, we study the average transversal complexity of road networks. By viewing road networks as multiscale-dispersed graphs, we show that a random line will cross the edges of such a graph $O(\sqrt{n})$ times on average. In addition, we provide by empirical evidence from experiments on the road networks of the fifty states of United States and the District of Columbia that this bound holds in practice and has a small constant factor. Combining this result with data structuring techniques from computational geometry, allows us to show that we can then do point location and ray-shooting navigational queries with respect to road networks in $O(\sqrt{n}\log n)$ expected time. Finally, we provide empirical justification for this claim as well.

**ACM Categories:** F.2 ANALYSIS OF ALGORITHMS AND PROBLEM COMPLEXITY; F.2.2 Nonnumerical Algorithms and Problems (Geometrical problems and computations); G.2.2 Graph Theory (Graph algorithms); G.2.2 Graph Theory (Network problems); H.2.8 Database Applications (Spatial databases and GIS)

# 1   Introduction

Road networks are a core topic of study in Geographic Information Systems (GIS), and they have been studied from many perspectives, including views of road networks as types of social networks, and geometric graphs.

## 1.1 Geometric Graphs

In this paper, we continue the study of road networks as geometric graphs. A *geometric graph* [1, 12, 22–24, 31] is a graph $G = (V, E)$, such that each vertex $v$ in $V$ is associated with a unique point $p(v)$ in the plane and each edge $e = (v, w)$ is associated with a curve segment that has $p(v)$ and $p(w)$ as its respective endpoints. In viewing a road network as a geometric graph, we create a vertex for every road intersection, major jog, and highway on-ramp and off-ramp. We then connect each pair of such vertices that have a road segment that joins them, and we typically view the curve associated with this pair as a straight line segment.

Typical applications of road networks include scenarios where road networks are queried to determine driving directions between two locations or to plan transportation flows. Of course, in these applications, we are interested only in movements that travel exclusively along the edges of the network. That is, in these applications, the geometry of a road network takes a back seat to the road network's graph structure in these applications, with edge length being more important than the orientation, angles, and geometric proximity of the different edges. This is not the only way to use a road network, however.

## 1.2 Going Off Road

In this paper, we study uses of a road network that involve movements that go "off road," but nevertheless need to interact with the components of that road network. For example, we may wish to maintain a locational awareness with respect to the roads in an environment for a flying autonomous vehicle as it flies over a terrain with an embedded road network while monitoring traffic flows and/or airborne pollutants. In such applications, we may need to monitor motions that involve transversals, that is straight lines or line segments that cut across the edges and open spaces in a road network rather than being restricted only to follow along the edges of the road network.

Some natural algorithmic questions that immediately arise, with respect to a given $n$-vertex road network, $G = (V, E)$, in such applications, include the following:

1. How many edges on average will an infinite straight line or ray cut across in $G$?

2. How many edges on average will a straight line segment cut across in $G$?

3. Can we design a simple data structure that can efficiently report, for any given line or line segment, $\ell$, the edges in $G$ that are cut by $\ell$?

4. Can we design a simple data structure that can efficiently report, for any given line segment, $\ell$, the face in $G$ that contains one of the endpoints of $\ell$ given the face containing the other endpoint of $\ell$?

Questions 1 and 2, therefore, deal with the *average transversal complexity* of road networks, that is, the average number of edges that are crossed by a random line or line segment. Questions 3 and 4 are, of course, strongly related to these first two questions, since the time complexity of such queries depend on the number of edges in $G$ that are crossed by the query segment.

Such questions are easily motivated and are the main subjects of this paper. We could imagine, for example, that we are planning a flight path for a flying autonomous vehicle and we would like to compute the edges of $G$ that this path is expected to cross. We illustrate some example line transversals for San Francisco in Figure 1 and some example line transversals for New York in Figure 2.

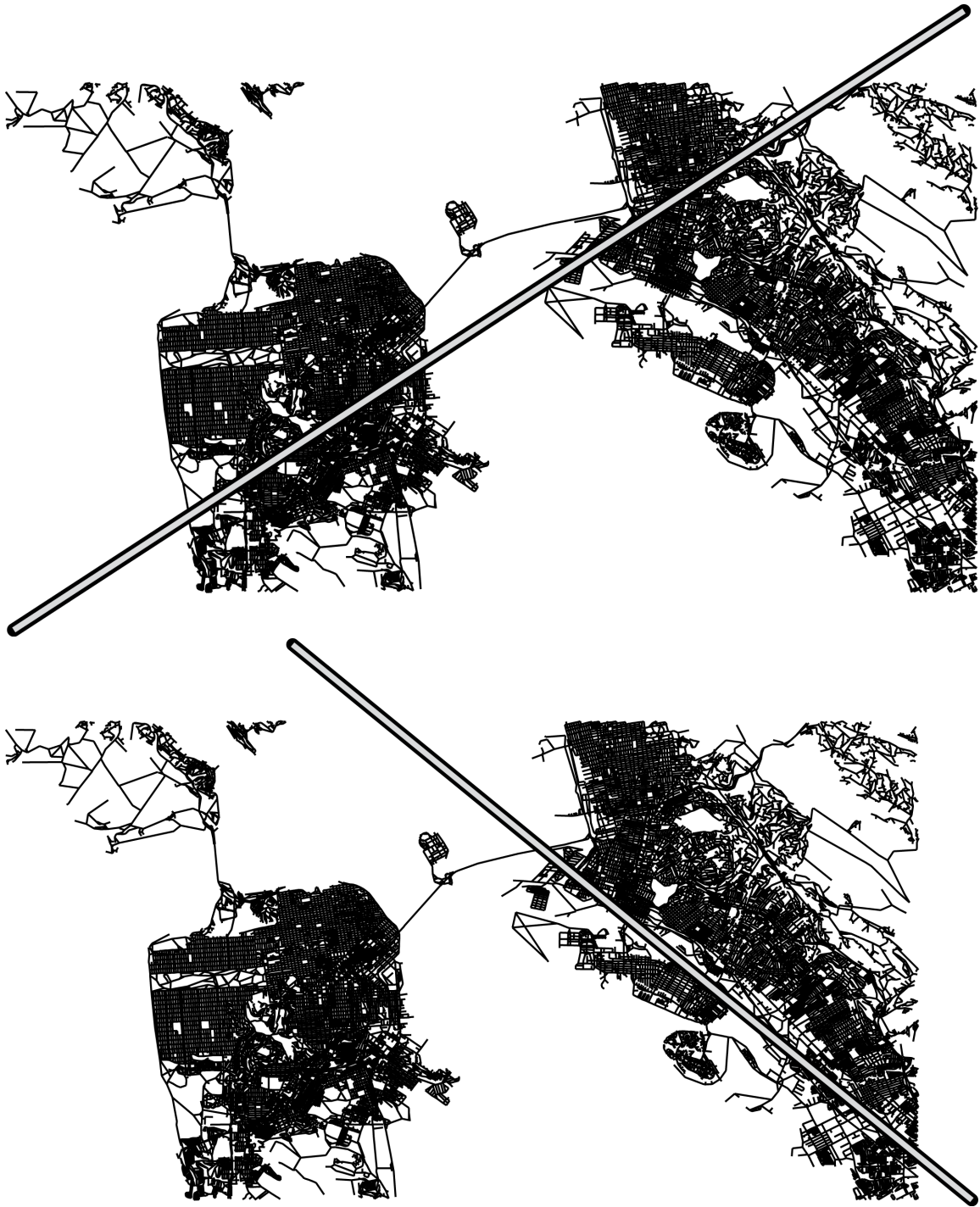

Figure 1: Two line transversals for San Francisco, using Tiger/Line data, which illustrate how significantly different transversal complexities are possible even in a big city.

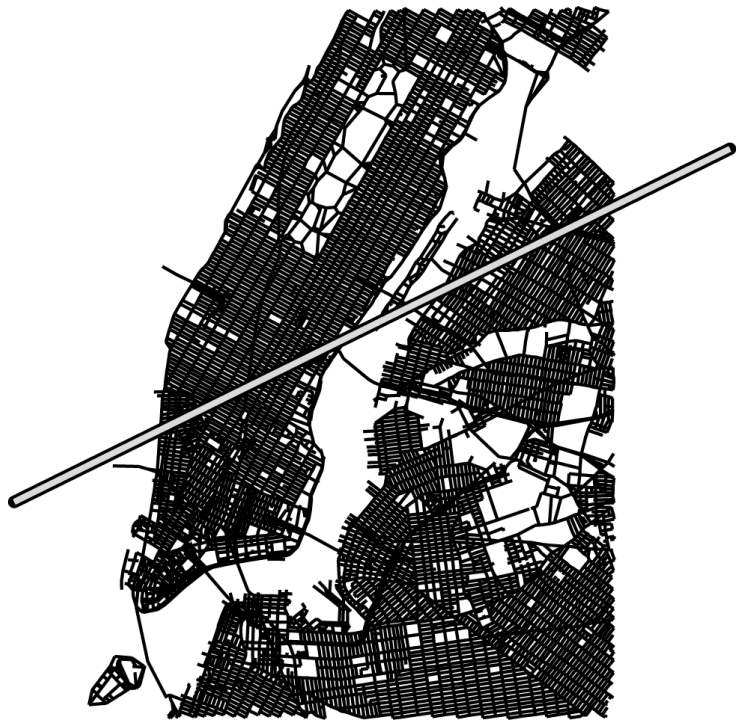

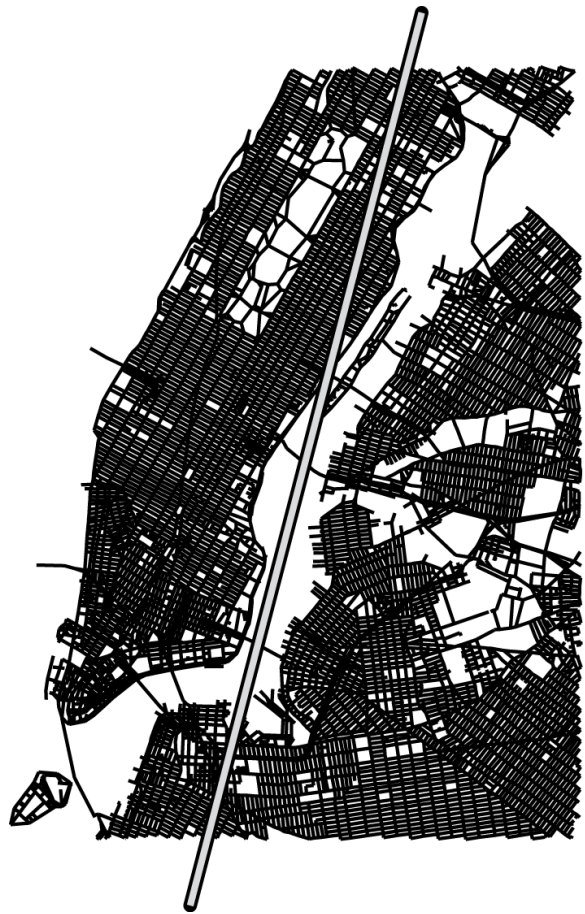

Figure 2: Two line transversals for New York, using Tiger/Line data, which also illustrate the different transversal complexities that are possible even in a big city, which motivates the need for an analysis based on average transversal complexity.

## 1.3 Related Work

There is considerable amount of prior related work on road networks in the transportation literature (e.g., see [3]) and in algorithms community, as well (e.g., see [13, 17, 18, 28, 30, 34] and the program for the Ninth DIMACS Implementation Challenge[1]).

The term "geometric graph" was popularized by Alon and Erdös [1] in 1989, although such structures were studied long before then (e.g., see [11, 19, 32, 33]). They define a geometric graph to be a graph $G = (V, E)$ embedded on a planar surface such that the vertices are distinct points in the plane and the edges are straight line segments joining pairs of those points. Examples of such geometric graphs include the natural abstractions of road networks, as well as planar convex hulls, Voronoi diagrams, Delaunay triangulations, line arrangements, visibility graphs, polygon triangulations, which collectively form the backbone of the topics in computational geometry (e.g., see [5, 8, 21, 27]). Thus, it should not be surprising that there is a developing modern interest in geometric graphs, in both mathematics and computer science (e.g., see [12, 22–24, 31]).

Eppstein and Goodrich [9] initiated a study of road networks in terms of their being viewed as special types of geometric graphs known as multiscale-dispersed graphs and they provided empirical evidence that real-world road networks are in fact instances of such graphs. Our study in this paper continues this viewpoint of road networks, but asks new types of questions, related to transversals, which were not studied by Eppstein and Goodrich, so include in this paper a review of this framework.

## 1.4 Our Results

In this paper, we study line and line segment transversals, addressing the questions mentioned above. Note, for instance, that it is possible to construct contrived examples of fictional road networks such that a random line will intersect $\Omega(n)$ edges on average in a road network with $n$ vertices (e.g., create a network consisting of $n/4$ nested squares all of roughly the same size, with their respective sides all close to one another). Nevertheless, by viewing road networks as types of geometric graphs known as multiscale-dispersed graphs, we show that any road network will be intersected at most $O(\sqrt{n})$ times on average by a random line, for a road network with $n$ vertices. This shows that, with respect to Questions 1 and 2, road networks have a "spreading" feature that avoids the type of transversal complexities that could be possible in the worst case. We also study the data structuring problems represented in Questions 3 and 4. We show that a simple augmentation of a standard representation for geometric graphs can be used to answer these types of queries in $O(n^{1/2} \log n)$ time on average for random line or line-segment queries.

## 2 Theoretical Analysis of the Transversal Complexity of Road Networks

We assume throughout this paper that the road networks of interest are connected graphs. In applications, such as in the presence of road networks on multiple islands, where the road network is disconnected, one can apply our techniques to each connected component separately.

---

[1]See http://dimacs.rutgers.edu/Workshops/Challenge9/

## 2.1 Multiscale-Dispersed Graphs

In this section, we review the framework and definitions introduced by Eppstein and Goodrich [9], for viewing road networks as multiscale-dispersed graphs, as this is the framework of geometric graphs we use for road networks in this paper.

Let $G = (V, E)$ be a road network with $n$ vertices, i.e., with $n = |V|$. For each vertex, $v$ in $G$, let $r(v)$ be the length of the longest edge incident on $v$, and let $D(v)$ be the closed disk centered at $v$ with radius $r(v)/2$. The set $\mathcal{D}$, of disks, $D(v)$, defined for all the vertices in $V$ in this way is known as the *natural disk neighborhood system* for $G$. Note that, by definition, the graph $G$ is a subgraph of the disk intersection graph defined by the disks in $\mathcal{D}$, where we create a vertex for each disk in $\mathcal{D}$ and connected each pair of vertices that correspond to intersecting disks. The *ply* of $\mathcal{D}$ is the maximum number of disks in $\mathcal{D}$ that are stabbed by any given point $p$ in the plane. We say that $G$ is a *multiscale-dispersed graph* if there is a set $T$ of at most $O(\sqrt{n})$ disks in $\mathcal{D}$, such that the disks in $\mathcal{D} - T$ have ply at most $O(1)$. The disks in $T$ are exceptions to the claim that all the disks in $\mathcal{D}$ for a constant-ply disk system. Thus, a graph $G$ is a multiscale-dispersed graph if its natural disk neighborhood system is an $O(\sqrt{n})$-exceptional $O(1)$-ply disk system [9].

## 2.2 Line and Line-Segment Transversal Complexities

In this subsection, we address the first two of our algorithmic questions from a theoretical perspective, to characterize the average number of edges in a road network that are crossed by a random line or line segment.

There is a canonical measure on lines in the plane, with the property that the length of any curve in the plane is proportional to the average number of crossings of the curve with a random line (e.g., see Miles [20] (Theorem 3) and others [6, 16]). This measure is invariant under rotation and translation of the plane, and (except for the lines through the origin, which have measure zero) can be generated by taking a random point on the plane, forming the vector from the origin to the point, and taking a line through the point and perpendicular to the vector. If one wants to restrict to the lines that intersect a given set $S$, enclose $S$ in a disk centered at the origin, restrict the randomly generated points to be uniform on the circle bounding that disk, and then filter out the lines that do not intersect $S$.

**Theorem 1:** *Consider a convex set $K$, and a set $S$ of $n$ disks contained in $K$, such that a random point from $K$ intersects $c = O(1)$ disks in expectation. Then a random line intersecting $K$ intersects $O(\sqrt{n})$ disks in $S$ in expectation.*

**Proof:** We begin by replacing $K$ by a minimal bounding disk and normalizing it to be the unit disk in the plane. A random line through the disk will intersect $K$ with constant probability, so this replacement doesn't change the asymptotic expected stabbing number by more than a constant factor. It does allow us to determine the constant of proportionality in the ratio between the length of a curve and the average crossing number of a random line with the curve, according to the canonical measure mentioned above, however. In particular, we know that the average crossing number of a line with the disk boundary is 2, since almost every line that intersects the disk crosses its boundary twice. Additionally, the length of the boundary of the unit disk, which contains all the disks in $S$, is $2\pi$. Thus, our normalization implies that the constant of proportionality with respect to the canonical measure mentioned above is $2/2\pi$; that is, the average crossing number of a random line that intersects the unit disk with a curve of length $L$ within the disk is $L/\pi$. Let $r_i$ denote the radius of the $i$-th disk in $K$, so that the length of the arrangement defined by the boundaries of all the disks in $K$ is

$$\sum_{i=1}^{n} 2\pi r_i.$$

Thus, the average crossing number of a random line with this arrangement is

$$\frac{(\sum_{i=1}^{n} 2\pi r_i)}{\pi},$$

which implies that the average number of disks that are intersected by a random line is

$$2\sum_{i=1}^{n} r_i.$$

Thus, the average crossing number is determined by the radii, $r_i$, of the disks in $S$. However, the values of the radii $r_i$ (and thus, indirectly, the average crossing number) are upper bounded by the assumption that the disks have low average ply. Specifically, a disk with radius $r_i$ has probability $r_i^2$ (the ratio of the area of the disk with that of $K$) of being intersected, so the expected intersection number is just the sum of these probabilities,

$$\sum_{i=1}^{n} r_i^2.$$

However, we have assumed that the expected intersection number is exactly $c$. Therefore,

$$\sum_{i=1}^{n} r_i^2 \leq c.$$

That is, the maximum number of intersections with a random line is at most the maximum value of $\sum 2r_i$ achievable for a set of positive numbers $r_i$ satisfying the constraint that $\sum r_i^2 \leq c$. This maximum is achieved when all $r_i$ are equal to each other, and each is equal to $\sqrt{c/n}$. For this choice of $r_i$'s the value of $\sum r_i^2$ is

$$\sum_{i=1}^{n} \left(\sqrt{c/n}\right)^2 = c,$$

as required, and the value of $2\sum r_i$ is

$$2\sum_{i=1}^{n} \sqrt{c/n} = 2\sqrt{cn},$$

which is $O(\sqrt{n})$. ∎

This theorem implies the following corollary.

**Corollary 2:** *The disks in a disk system corresponding to a multiscale-dispersed graph are stabbed $O(\sqrt{n})$ times in expectation by a random line.*

**Proof:** In a multiscale-dispersed graph there are $O(\sqrt{n})$ exceptional disks that may or may not all be stabbed. Considering only the remaining disks, we have a disk system with $O(1)$-ply. We can therefore apply Theorem 1 to the remaining disks, to show that $O(\sqrt{n})$ of these disks are stabbed. Therefore, $O(\sqrt{n})$ of all of the disks will be stabbed by a random line. ∎

We also have the following.

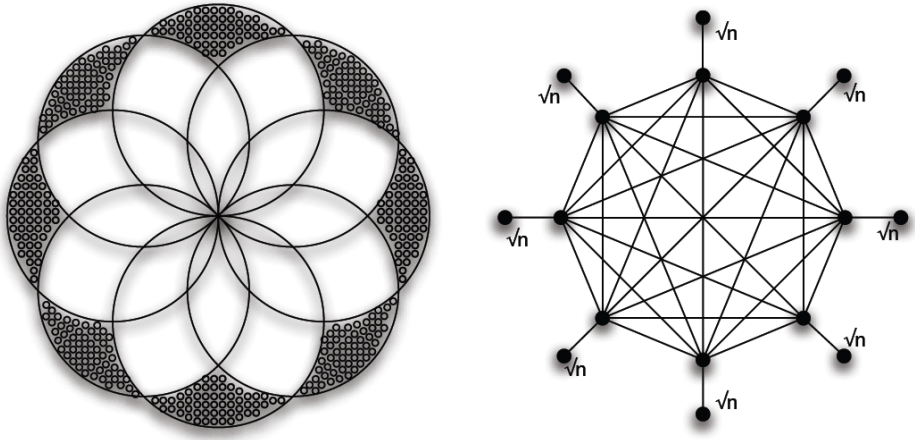

Figure 3: Non-bounded degree counter-example.

**Corollary 3:** *In a multiscale-dispersed graph with maximum vertex degree $\Delta$, a random line crosses $O(\Delta\sqrt{n})$ edges in expectation.*

**Proof:** If a random line crosses an edge, it crosses at least one of the disks corresponding to the edge endpoints, and we can charge the edge crossing to one of these two disk crossings. By Corollary 2, there are $O(\sqrt{n})$ disk crossings in expectation; each of these can be charged at most $\Delta$ times by nearby edge crossings. Therefore, the total expected number of edge crossings is $O(\Delta\sqrt{n})$. ∎

More simply stated, in a multiscale-dispersed graph in which the vertex degree is bounded by a constant, a random line crosses $O(\sqrt{n})$ edges in expectation.

Thus, we show that a theoretical analysis applied to Questions 1 and 2 from the introduction implies that the line or line-segment transversal complexity of a road network with $n$ vertices is $O(\sqrt{n})$.

### 2.3 Necessity of the Degree Bound

The assumption in the previous section that the maximum vertex degree of a multiscale-dispersed graph is bounded does not cause any difficulty in applying our results to real-world road networks, as all such networks have low vertex degree. However, is this assumption necessary even in theory, or is it possible to bound the number of crossings with a random line independently of the vertex degree?

As we now observe, the answer is that the dependence on degree is necessary, as there exist non-bounded-degree multiscale-dispersed graphs where the average crossing number of a random line with the graph is $\Omega(n)$. For instance, make $O(\sqrt{n})$ unit disks centered on a circle of radius $1-\varepsilon$ (so they all contain the origin), and $O(n)$ tiny disks disjoint from each other and each contained in one of the large disks. (See Figure 3.) The resulting intersection graph is a multiscale-dispersed graph in the form of a clique of $O(\sqrt{n})$ vertices together with $O(n)$ degree-one vertices. The graph has $\Omega(n)$ edges connecting the centers of the small disks to the centers of the large disks, and each of these edges has length $\Omega(1)$, from which it follows from the canonical measure on lines that a random line cuts $\Omega(n)$ clique edges.

## 3 Experimental Transversal Results on Road Networks

In this section, we examine Questions 1 and 2 from an empirical perspective, studying experimental results on stabbing road map data with random lines. Our analysis uses the U.S. TIGER/Line road network database, as provided by the Ninth DIMACS Implementation Challenge, which is comprised of over 24 million vertices and 29 million edges.

### 3.1 Experimental Results for Lines

It is possible to conceptualize why road maps seem to have $O(\sqrt{n})$ crossings if we consider the design of common roadway, with roads in parallel and intersections forming a grid. Now imagine a large highway cut through an otherwise nice map. This line superimposed on the aforementioned grid will result in the $O(\sqrt{n})$ crossings that our experiments return.

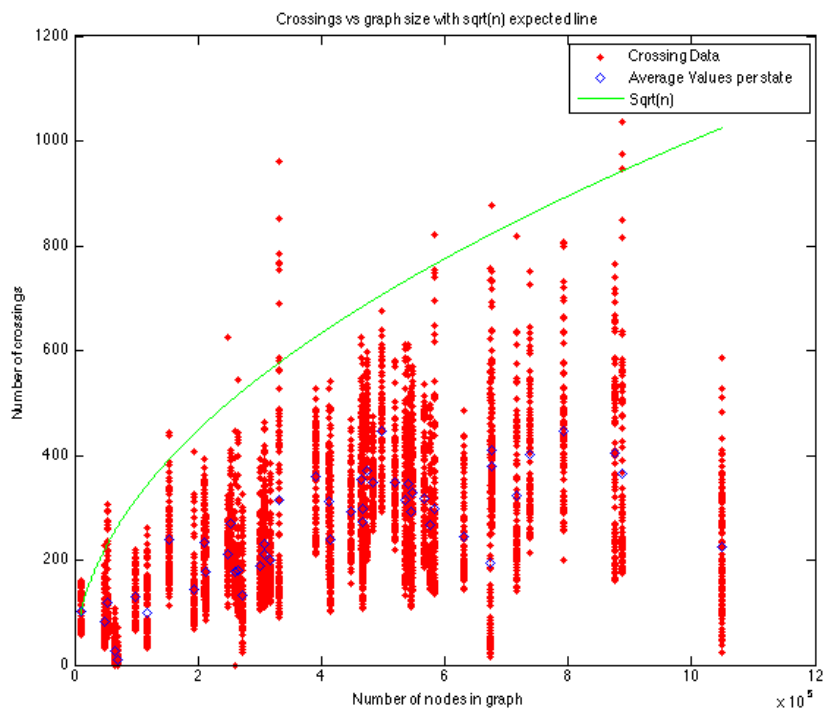

Figure 4: Road network crossings.

If we examine Figure 4, we can see that although most often the number of crossings is below the $\sqrt{n}$ trend line, the numbers of crossings of a random line with the edges of the tested graphs do tend to increase with the size of the graph, and the largest of these numbers appear to be proportional to the root of the number of nodes in the road network.

It is plausible to explain that while the nature of the roadways as mentioned above does dictate a theoretical bound of $O(\sqrt{n})$ crossings, the reason the empirical results usually lie below $\sqrt{n}$ involves the relative scales of the geographic regions represented within the road maps being tested. The grid structures that could generate large crossing numbers are confined to large cities, which are relatively small compared to the sizes of most states and are geographically dispersed. Therefore, we would probably have to restrict our attention to cities to more closely approach the $O(\sqrt{n})$ upper-bound complexity of our theoretical analysis.

## 3.2 Experimental Results for Line Segments

Although the canonical measure on lines provides an elegant theory of expected crossings numbers, we are also interested in studying crossing numbers for line segments. We next study line segment transversal complexity for road networks from an empirical perspective.

Figure 5 shows that for our road network data the number of edges crossed by a random line segment. Note that the number of segments crossed is low. We performed this experiment as follows. First two points were randomly generated with the bounding disk of the graph. We then determined how many graph edges cross the line segment connecting these two points. These numbers were then recorded. Interestingly, the number of edges crossed by a random line segment lies well below the $O(\sqrt{n})$ bound in these experiments.

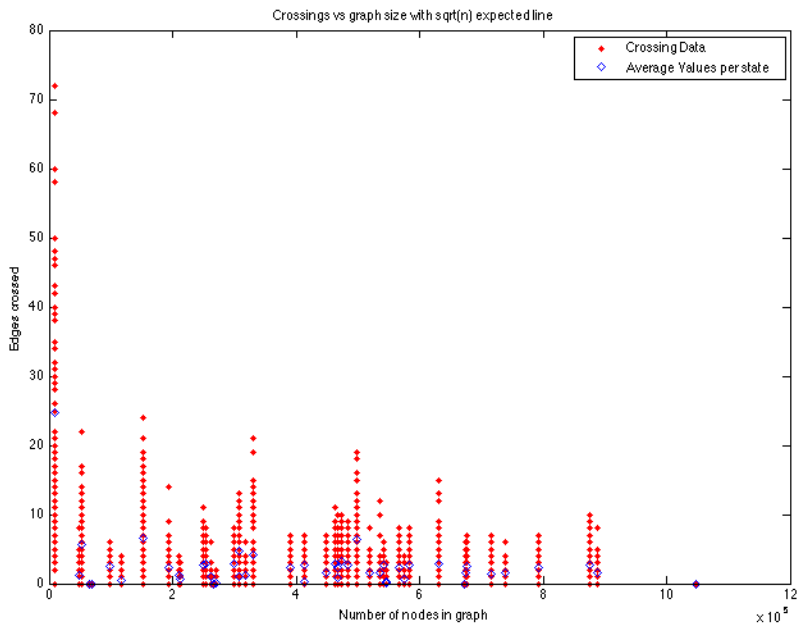

Figure 5: Point navigation numerical crossings.

## 4 Ray-Shooting Navigation

The next problems we address is navigation from one point contained within a road network environment to another point via ray-shooting, addressing the algorithmic Questions 3 and 4 mentioned in the introduction. We describe in this section a data structure that allows us to traverse across the edges of a road network efficiently.

For any query ray $\vec{r}$, we assume that we have located the starting point of $\vec{r}$ to a face or edge of the embedded graph defined by the road network. This starting point could have been the end point of a previous query or it could be the starting point on an edge that is the beginning of a flight path.

## 4.1 The Ray-Shooting Data Structure

We construct our data structure for a road network $G = (V, E)$ as follows. We use the fact that road networks are geometric graphs having $O(\sqrt{n})$ edge crossings to construct the arrangement of the segments in $G$ in $O(n)$ time, using the algorithm of Eppstein *et al.* [10]. In this embedding, $G'$, each vertex in $G$ is included as a vertex in $G'$, but each edge crossing between two intersecting edges in $G$ is also included as a vertex in $G'$ (even though it is not explicitly represented in $G$). We connect each vertex in $G'$ with the longest segment from an edge in $G$ that includes no vertices of $G'$ in its interior. Thus, we get a planar embedding.

In addition, let us assume that we are using the widely-used and well-known doubly-connected edge list (DCEL) representation (e.g., see [5,21,27]). for the graph $G'$. This choice implies that each edge object stores pointers to the two vertex objects representing its end vertices, each vertex object stores a pointer to an implicitly defined linked list of incident edges, stored in counter-clockwise order. Likewise, we represent each edge as a double edge, so that each edge is assumed to have a "left" side and a "right" side, and we link these two sides by reciprocal pointers between their respect object representations. This representation gives us an embedding of $G$ in the plane so that each face is equivalent to a simple polygon, since we assume that $G$ is connected and we can view each edge as having two sides (which would be listed separately in a traversal of a face having "dangling" edges in its interior).

For each face of $G'$, then, we construct a geodesic triangulation, using the method of Chazelle *et al.* [2], so that ray shooting in any face can be done in $O(\log n)$ time to determine the first boundary edge hit by any query ray. This involves adding additional "dummy edges" inside each face, so that each interior face becomes a geodesic triangle, that is, a face formed by three concave polygonal chains. In addition, for each face, we store representations of each of its three chains in a binary search tree, as described by Chazelle *et al.* [2]. So this augmentation amounts to a modest extension to the DCEL representation of $G'$ itself. Moreover, this method allows for very simple ray traversals, which can be implemented simply by marching through the geodesic triangles of the geodesic triangulation, with their edge lists stored in weighted binary search trees. That is, when a ray hits an edge, we march to its other side, using the link to its "twin," and when we march into a face forming a geodesic triangle, we use the binary search trees for its other two chains to perform a simple search for the next edge in this face that our ray hits.

## 4.2 Performance Analysis

Chazelle *et al.* show that we can set up the binary search trees so that this approach to answering ray-shooting queries via simple traversals through faces and across edges results in a query time of $O(\log n)$ to find the next (real) edge hit by a query ray in a face of a simple polygon. Given our previous analysis, therefore, we can easily analyze the complexity of performing a search in this data structure. Let us consider our starting point somewhere in a face of $G$, augmented as described above, to a geodesic triangulation of $G'$. We have two cases for the location of the terminal point. The first possibility is that this point is within the same simple polygon formed by the edges of the graph as our start point. If so, we know from Chazelle *et al.* [2] that navigation to this point can be done in $O(\log n)$ time using the geodesic triangulation. The second case is that our terminal point lies outside our originating simple polygon. If so, the cost will be $O(\log n)$ to establish a new starting point in the adjacent polygon. We can then iterate this procedure. Since we have shown above that a random line will stab $O(\sqrt{n})$ edges, we know that our we will cross at most $O(\sqrt{n})$ edges, on average, with a random ray-shooting query, before reaching our terminal point. Since each polygon crossed requires $O(\log n)$ time to find the place of intersection on the other side, we can conclude that this procedure can be completed in $O(\sqrt{n} \log n)$ expected time. Thus, we have the following.

**Theorem 4:** *Given a connected road network $G$ of $n$ vertices, we can construct a data structure of size $O(n)$ that can answer a random ray-shooting query with respect to $G$ in $O(n^{1/2} \log n)$ expected time.*

**Proof:** We described our scheme assuming that we had already located the starting point of a ray-shooting query with respect to the geodesic triangulation. We can relax this assumption, however, simply by augmenting our geodesic triangulation with a point-location data structure [4, 7, 15, 25, 26, 29], which can locate the starting point of a ray in $O(\log n)$ time. Such an augmentation is not actually needed, however. Given a starting point, $p$, for a random ray, we can actually locate $p$ with respect to our geodesic triangulation by running our ray-shooting procedure to go from any random vertex in $G$ to $p$ using the same traversal process. This traversal will itself define a random line segment that terminates with $p$ and we can then traverse the given random ray that starts from $p$ using a similar method. Thus, a point location data structure augmentation is actually optional in this approach, from a performance standpoint. ∎

Moreover, from work completed by Goodrich and Tamassia [14], a relaxation to $O(\sqrt{n} \log^2 n)$ time allows us to perform the ray-shooting queries in a dynamically changing road network, where roads can be periodically inserted and/or deleted.

## 5 Conclusion

In this paper, we have studied the transversal complexity of road networks, both from theoretical and empirical standpoints. Our theoretical analysis shows that the geometric graphs that share the same properties as road networks will have $O(\sqrt{n})$ edge crossings, on average, with a random line or line segment, which is significantly better than the worst-case, which is $\Theta(n)$. Our experimental analysis shows that even our theoretical analysis is likely an over-estimate, especially for line segment queries, as these are often bounded by a constant. In addition, we have given a simple data structure for answering ray-shooting queries in road networks, which is shown to be fairly efficient, based on our analysis. Our data structure has linear space and amounts to little more than a simple augmentation of the doubly-connected edge list (DCEL) representation of a planarization of the road network itself. It answers ray-shooting transversal queries in time that is proportional to the transversal complexity times a logarithmic factor.

### Acknowledgments

This research was supported in part by the U.S. National Science Foundation, under grants 0713046, 0830403, and 0847968, and by an Office of Naval Research: Multidisciplinary University Research Initiative (MURI) Award, number N00014-08-1-1015.